\documentstyle[12pt]{article}

\begin{document}

\vspace*{0.5cm}

\begin{center}
{\large {\bf Nuclear dependence asymmetries in direct photon production
\footnote{Supported in part by National Natural Science Foundation of China}
 }      }
\end{center}

\vspace*{1.0cm}
Hong Shen$^{a,b,}$\footnote{email address: songtc@public1.tpt.tj.cn} and Ma Luo$^{a,b}$ \\

   $^{a}$CCAST(World Lab.), P.O.Box 8730, Beijing 100080, China

   $^{b}$Department of Physics, Nankai University, Tianjin 300071, China
 \\

\vspace*{0.5cm}

\begin{abstract}
   We study the nuclear dependences of high-$p_T$
   jet cross sections in  one photon and one jet production 
   in proton-nucleus collisions. We find that there exist asymmetries 
   between the outgoing jets and photons.
   A convincing reason responsible for those asymmetries are demonstrated
   in perturbative QCD. Significant nuclear enhancements 
   are also found in the inclusive jet cross sections.   
\end{abstract}

 
Anomalous nuclear dependences are observed in hadron-nucleus
scatterings~\cite{a1}, as well as in some extensive experimental 
data~\cite{a2,a3}. It is shown that the inclusive cross sections 
for single high-$p_T$ particles rise rather faster than 
proportional to the atomic number $A$ of the nuclear target,
while a linear $A$ dependence is naively expected in single scattering model.
This behavior is often described as due to multiple scatterings
of the partons in the nuclear matter, primarily due to double scatterings~\cite{a4,a5,a6,a7}.

In ref.[5], M. Luo, J. Qiu and G. Sterman showed that the anomalous nuclear
enhancements in deeply inelastic scatterings and photoproductions
could be described naturally in perturbative QCD, in term of 
a higher twist formalism. They presented the details of the
use of factorization at higher twist to describe multiple scattering 
in nuclei. The contribution from multiple scattering was factorized
into a calculable short-distance partonic part multiplied by a
multiparton correlation
function. The double scattering must have one hard scattering  
to produce the high-$p_T$ observables
along with another scattering
which, in principle, could be either soft or hard. 
It was also pointed that the one hard plus one soft scatterings
would dominate the nuclear dependences.
Moreover the soft scattering could occur either before or after the hard scattering (referred to below as initial-state or final-state multiple scattering). The initial-state multiple scatterings, however, are absent 
in deeply inelastic scattering and photoproduction because of no direct
interaction between photons and gluons, therefore 
only the final-state multiple scatterings were discussed in ref.[5].
 
In order to study the contributions from the initial-state multiple
scatterings, X. Guo and J. Qiu extended the higher twist perturbative QCD
treatment to high-$p_T$ direct photon 
productions in hadron-nucleus scatterings~\cite{a6}. The direct photon 
productions provide excellent tests of initial-state multiple scatterings,
since the final-state multiple scatterings are  simply ruled out due to
the fact that none of gluons from the nuclear target could interact 
with the final photon. They calculated the 
nuclear dependences of the direct photon inclusive cross sections
in hadron-nucleus collisions. Small nuclear enhancements contributed
from initial-state multiple scatterings were obtained and discussed in ref.[6],
which were also compared with the data from Fermilab
experiment E706~\cite{a3}.

In this paper, we work on the same reaction ($P+A \rightarrow\gamma+jet+X$). 
We concentrate, however, on the inclusive cross section of the final jet
instead of the direct photon. We demonstrate that the final jet receives 
contributions from both the
initial-state and final-state multiple scatterings, 
while the direct photon receives contribution from the initial-state 
multiple scattering only,
which results in the asymmetry between the distributions of the final jet
and the direct photon.
We calculate the final jet inclusive cross section, and reproduce
the direct photon inclusive cross section which has been derived in ref.[6]. 
In order to compare with ref.[5] and [6], we adopt the same multiple
correlation functions and similar notations for the formalism in the following discussion. 

The process we study here is that one photon and one jet with 
high-$p_T$ produced in proton-nucleus collision,
$P+A \rightarrow\gamma+jet+X$. 
The inclusive cross section for either the final jet or the direct photon 
can be expressed as
a sum of contributions from single and double scatterings, while
contributions from triple and even higher multiple scatterings are ignored,
\begin{equation}
  E_l\frac{d\sigma      _{i}(l)} {d^3l}
 =E_l\frac{d\sigma^{(S)}_{i}(l)} {d^3l} 
 +E_l\frac{d\sigma^{(D)}_{i}(l)} {d^3l},     	
\end{equation}
where we use the index $i$ to identify the inclusive cross section 
in which the $i$-particle ($i=jet$ or $\gamma$) is detected with 
momentum $l$, the superscripts ($S$) and ($D$) represent 
the single and double scatterings, respectively. 
In this paper we emphasize the difference between the inclusive 
cross sections
$E_l\frac{d\sigma_{jet}(l)} {d^3l}$
and 
$E_l\frac{d\sigma_{\gamma}(l)} {d^3l}$
, and its anomalous nuclear enhancement.

The single scattering cross section is proportional to $A$ as
\begin{eqnarray}
E_l\frac{d\sigma^{(S)}_{i}(l)} {d^3l} &=&
A\displaystyle\sum_{a,b} \int dx^{\prime} f_{a/P}(x^{\prime} ) 
                         \int dx f_{b/N} (x) 
    \delta \left(x-\frac{-x^{\prime} t}{x^{\prime}s+u}\right) \\  \nonumber
 & & \times \alpha_{em}\alpha_s\left(\frac{1}{\hat{s}}\right)
     \left(\frac{1}{x^{\prime}s+u}\right)
     |\bar{M}_{ab\rightarrow\gamma jet}|^2,
\end{eqnarray}
where $f_{a/P}(x^{\prime})$ and $f_{b/N}(x)$ are regular parton 
distribution functions with the $P$ incident proton and the $N$ 
effective nucleon in the nuclear target,
the $a$ and $b$ run over all gluon, quark,and antiquark flavors 
by $\displaystyle\sum_{a,b}$. The matrix elements of the subprocesses 
in eq.(2) are given by
\begin{eqnarray}
{|\bar{M}_{gq\rightarrow\gamma q}|}^2 &=&
e^2_q\;\left(\frac{1}{6}\right)\;2\;
\left(\frac{-\hat{u}}{\hat{s}}+\frac{\hat{s}}{-\hat{u}}\right),     
               \\
{|\bar{M}_{qg\rightarrow\gamma q}|}^2 &=&
e^2_q\;\left(\frac{1}{6}\right)\;2\;
\left(\frac{-\hat{t}}{\hat{s}}+\frac{\hat{s}}{-\hat{t}}\right),
               \\
{|\bar{M}_{q\bar{q}\rightarrow\gamma g}|}^2 &=&
e^2_q\;\left(\frac{4}{9}\right)\;2\;
\left(\frac{\hat{u}}{\hat{t}}+\frac{\hat{t}}{\hat{u}}\right),
\end{eqnarray}
where $e_q$ is the fractional quark charge, and the
invariants $\hat{s}$, $\hat{t}$, and $\hat{u}$ are 
\begin{equation}
 \hat{s}=x^{\prime}x(p^{\prime}+p)^2,\hspace{1cm}  
 \hat{t}=x^{\prime} (p^{\prime}-l)^2,\hspace{1cm}
 \hat{u}=x          (p-l)^2,
\end{equation}
with the $p^{\prime}$ and $p$ are the momenta of the incident proton and
the effective nucleon in nuclear target, respectively.
For the single scattering term in eq.(1) we simply have
\begin{equation}
  E_l\frac{d\sigma^{(S)}_{jet   }(l)} {d^3l}
 =E_l\frac{d\sigma^{(S)}_{\gamma}(l)} {d^3l},     	
\end{equation}
and the asymmetry of the 
$ E_l \frac{d\sigma_{i}(l)}{d^3l} $ ($i=jet$ or $\gamma$)
majorly comes from their double scattering terms, 
$ E_l \frac{d\sigma^{(D)}_{i}(l)}{d^3l} $.

The double scattering cross section of the final jet,
$ E_l \frac{d\sigma^{(D)}_{jet}(l)}{d^3l} $,
is contributed from 
both final-state and initial-state interactions, 
\begin{equation}
  E_l\frac{d\sigma^{(D)}_{jet}(l)} {d^3l}
 =E_l\frac{d\sigma^{(D-f)}_{jet}(l)} {d^3l} 
 +E_l\frac{d\sigma^{(D-i)}_{jet}(l)} {d^3l},     	
\end{equation}
while the double scattering cross section of the direct photon receives contribution from the initial-state interaction only~\cite{a6},
\begin{equation}
  E_l\frac{d\sigma^{(D)}_{\gamma}(l)} {d^3l}
 =E_l\frac{d\sigma^{(D-i)}_{\gamma}(l)} {d^3l}.
\end{equation}
Since the initial-state interaction occurs before the hard scattering, 
which produce the final jet and the direct photon equally,
the initial-state interaction term in the 
double scattering cross section of jet,
$ E_l \frac{d\sigma^{(D-i)}_{jet}(l)}{d^3l} $,
has the same formalism as that of the direct photon,
$ E_l \frac{d\sigma^{(D-i)}_{\gamma}(l)}{d^3l} $,
which has been given in ref.[6] in detail.
As a result, in the following discussion we only need to work on 
the derivation of the term
$ E_l \frac{d\sigma^{(D-f)}_{jet}(l)}{d^3l} $.

A regular factorization procedure can separate the incident proton
part out, and turn the jet cross section to 
\begin{equation}
E_l\frac{d\sigma^{(D-f)}_{jet} (l) }{d^3l}=
\displaystyle\sum_{a}\int dx^{\prime} f_{a/P}(x^{\prime} )
\;\;
E_l\frac{d\sigma^{(D-f)}_{aA\rightarrow jet}(x^{\prime},l)}
{d^3l}. 
\end{equation}
In this topic the cross sections 
$E_l\frac{d\sigma^{(D-f)}_{aA\rightarrow jet}(x^{\prime},l)}{d^3l}$
can be classified in three groups shown in fig.1(1)-fig.1(3).
As a result of collinear expansion~\cite{a5,a6},
each individual sort of these cross sections in turn can be 
decoupled to a product of a hard process cross section multiplied by 
a corresponding four operator matrix element, such as for fig.1(1),
\begin{eqnarray}
E_l\frac{d\sigma^{(D-f)(1)}_{aA\rightarrow jet}
          (x^{\prime},l) } {d^3l}
&=&-\frac{1}{16x^{\prime}s}
\int\frac{dy^{-}_1}{2\pi}\frac{dy^{-}}{2\pi}\frac{dy^{-}_2}{2\pi}
\langle A|\bar{\psi}_q(0)\gamma^{+}
F_{+\bot}(y^{-}_2)F_{+}^{\bot}(y^{-})
\psi_q(y^{-}_1)|A\rangle
        \\ \nonumber
& & \times
\frac{\partial^2}{\partial k_{\bot}\partial k^{\bot}}  
\left\{\int dx dx_k dx_{k^{\prime}}
      e^{ixp^{+}y^{-}_1} e^{ix_kp^{+}y^{-}} 
      e^{-i(x_k-x_{k^{\prime}}) p^{+}y^{-}_2}
 \right.     \\ \nonumber
& & \times
 \left. \left. Tr\left[H_{\alpha\beta}^{(1)}(x^{\prime},x,x_k,x_{k^{\prime}},k_\bot,l)
            \gamma\cdot p \;\; p^\alpha p^\beta \right]
\right\} \right|_{k_\bot=0},
\end{eqnarray}
where the partonic hard scattering function
$H_{\alpha\beta}^{(1)}(x^{\prime},x,x_k,x_{k^{\prime}},k_\bot,l)$ 
is specified by the lower part in fig.1(1); similarly, for fig.1(2),
\begin{eqnarray}
E_l\frac{d\sigma^{(D-f)(2)}_{aA\rightarrow jet}
          (x^{\prime},l) } {d^3l}
&=&-\frac{1}{16x^{\prime}s}
\int\frac{dy^{-}_1}{2\pi}\frac{dy^{-}}{2\pi}\frac{dy^{-}_2}{2\pi}
\langle A|A^\bot(0)
F_{+\bot}(y^{-}_2)F_{+}^{\bot}(y^{-})
A_\bot(y^{-}_1)|A\rangle
        \\ \nonumber
& & \times
\frac{\partial^2}{\partial k_{\bot}\partial k^{\bot}}  
\left\{\int dx dx_k dx_{k^{\prime}}
      e^{ixp^{+}y^{-}_1} e^{ix_kp^{+}y^{-}} 
      e^{-i(x_k-x_{k^{\prime}}) p^{+}y^{-}_2}
 \right.     \\ \nonumber
& & \times
 \left. \left. Tr\left[H_{\alpha\beta}^{(2)}(x^{\prime},x,x_k,x_{k^{\prime}},k_\bot,l)
            p^\alpha p^\beta \right]
\right\} \right|_{k_\bot=0};
\end{eqnarray}
and for fig.1(3),
\begin{eqnarray}
E_l\frac{d\sigma^{(D-f)(3)}_{aA\rightarrow jet}
          (x^{\prime},l) } {d^3l}
&=&-\frac{1}{16x^{\prime}s}
\int\frac{dy^{-}_1}{2\pi}\frac{dy^{-}}{2\pi}\frac{dy^{-}_2}{2\pi}
\langle A|\bar{\psi}_q(0)\gamma^{+}
F_{+\bot}(y^{-}_2)F_{+}^{\bot}(y^{-})
\psi_q(y^{-}_1)|A\rangle
        \\ \nonumber
& & \times
\frac{\partial^2}{\partial k_{\bot}\partial k^{\bot}}  
\left\{\int dx dx_k dx_{k^{\prime}}
      e^{ixp^{+}y^{-}_1} e^{ix_kp^{+}y^{-}} 
      e^{-i(x_k-x_{k^{\prime}}) p^{+}y^{-}_2}
 \right.     \\ \nonumber
& & \times
 \left. \left. Tr\left[H_{\alpha\beta}^{(3)}(x^{\prime},x,x_k,x_{k^{\prime}},k_\bot,l)
            \gamma\cdot p \;\; p^\alpha p^\beta \right]
\right\} \right|_{k_\bot=0}.
\end{eqnarray}

Apparently, two more extra gluons occur in the double scattering as compared with single scattering case. they go through the four operator matrix element to form a virtual loop, so that the variables $x_k$ and $x_{k^\prime}$ carried by these extra gluons, in principle, can run over from $-\infty$ to $+\infty$.
In order to do the integrals $\int_{-\infty}^{+\infty} dx_k dx_{k^{\prime}}$
we extend the $x_k$ and $x_{k^\prime}$ to their complex planes,
then by using the contour integral technic carry the integrals 
$\int_{-\infty}^{+\infty} dx_k dx_{k^{\prime}}$ out~\cite{a5}.
As a result, the $x_k$ and $x_{k^\prime}$ are fixed at zero.
The vanish $x_k$ and $x_{k^\prime}$ in turn set the relation,
\begin{equation}
      e^{ix_kp^{+}y^{-}} 
      e^{-i(x_k-x_{k^{\prime}}) p^{+}y^{-}_2}
      =1.
\end{equation}
In this way, we eventually factorize the four operator matrix elements out of
the convolution along with the partonic hard scattering functions
$H_{\alpha\beta}^{(i)}$ with $i=1-3$.
These factorized four  operator matrix elements are responsible for
the twist-4 partonic distribution functions, they are read as:
for fig.1(1) and fig.1(3) we have
\begin{equation}
 T_q(x,A)=\int \frac{dy^{-}_1}{2\pi} dy^{-} \frac{dy^{-}_2}{2\pi}
                 e^{ixp^{+}y^{-}_1}
Tr\left[\langle A|\bar{\psi}_q(0)\frac{\gamma^{+}}{2}
F_{+\bot}(y^{-}_2)F_{+}^{\bot}(y^{-})
\psi_q(y^{-}_1)|A\rangle\right]
\theta (y^{-}-y^{-}_1)\theta(y^{-}_2),
\end{equation}
for fig.1(2),
\begin{equation}
 T_g(x,A)=\int \frac{xp^+dy^{-}_1}{2\pi} dy^{-}\frac{dy^{-}_2}{2\pi}
                 e^{ixp^{+}y^{-}_1}
\langle A|A^{\bot}(0)F_{+\bot}(y^{-}_2)F_{+}^{\bot}(y^{-})A_{\bot}(y^{-}_1)
|A\rangle
\theta (y^{-}-y^{-}_1)\theta(y^{-}_2),
\end{equation}
where the factor $\theta (y^{-}-y^{-}_1)\theta(y^{-}_2)$ results from the 
contour integrals~\cite{a5,a6}. Since the factor 
$\theta (y^{-}-y^{-}_1)\theta(y^{-}_2)$ can not confines the integral 
$\int_{-\infty}^{+\infty} dy^{-}$ going to infinity, extra contributions
proportional to the nuclear size, or $A^{1/3}$, occur in these double scattering
processes, which physically result in the anomalous nuclear 
enhancements~\cite{a5,a6}.

Alongside figuring these long distance partonic distributions out, we obtain
the double scattering cross sections,
\newpage
\begin{eqnarray}
E_l\frac{d\sigma^{(D-f)(1)}_{aA\rightarrow jet}
          (x^{\prime},l) } {d^3l}
&=&\alpha_{em}(4\pi\alpha_s)^2\;e^2_q\;
\left(\frac{1}{2x^{\prime}s}\right)\;
\left(\frac{1}{x^{\prime}s+u}\right)\;
           2 \; H_{q} \\ \nonumber
& &     \times
\left\{ \left[\frac{\partial^2}{\partial x^2}
        \left(\frac{T_q(x,A)}{x}\right)\right]
        \left(\frac{x^{\prime}s\;l_T}{u(x^{\prime}s+u)}\right)^2
     +  \left[\frac{\partial}{\partial x}
        \left(\frac{T_q(x,A)}{x}\right)\right]
        \left(\frac{-x^{\prime}s}{u(x^{\prime}s+u)}\right)
\right\},
\end{eqnarray}
\begin{eqnarray}
E_l\frac{d\sigma^{(D-f)(2)}_{aA\rightarrow jet}
          (x^{\prime},l) } {d^3l}
&=&\alpha_{em}(4\pi\alpha_s)^2\;e^2_q\;
\left(\frac{1}{2x^{\prime}s}\right)\;
\left(\frac{1}{x^{\prime}s+u}\right)\;
           2 \; H_g \\ \nonumber
& &   \times
\left\{ \left[\frac{\partial^2}{\partial x^2}
        \left(\frac{T_g(x,A)}{x}\right)\right]
        \left(\frac{x^{\prime}s\;l_T}{u(x^{\prime}s+u)}\right)^2
     +  \left[\frac{\partial}{\partial x}
        \left(\frac{T_g(x,A)}{x}\right)\right]
        \left(\frac{-x^{\prime}s}{u(x^{\prime}s+u)}\right)
\right\},
\end{eqnarray}
\begin{eqnarray}
E_l\frac{d\sigma^{(D-f)(3)}_{aA\rightarrow jet}
          (x^{\prime},l) } {d^3l}
&=&\alpha_{em}(4\pi\alpha_s)^2\;e^2_q\;
\left(\frac{1}{2x^{\prime}s}\right)\;
\left(\frac{1}{x^{\prime}s+u}\right)\;
           2 \; H_{q\bar{q}} \\ \nonumber
& & \times
\left\{ \left[\frac{\partial^2}{\partial x^2}
        \left(\frac{T_q(x,A)}{x}\right)\right]
        \left(\frac{x^{\prime}s\;l_T}{u(x^{\prime}s+u)}\right)^2
     +  \left[\frac{\partial}{\partial x}
        \left(\frac{T_q(x,A)}{x}\right)\right]
        \left(\frac{-x^{\prime}s}{u(x^{\prime}s+u)}\right)
\right\},
\end{eqnarray}
where the partonic parts $H_{q}$, $H_{g}$ and $H_{q\bar{q}}$ are given by
\begin{equation}
 H_{q}=\left(\frac{1}{36}\right)
       \left(\frac{x^{\prime}s}{-u}+\frac{-u}{x^{\prime}s}\right),
\end{equation}
\begin{equation}
 H_{g}=\left(\frac{1}{36}\right)
       \left(\frac{x^{\prime}s}{x^{\prime}s+u}
            +\frac{x^{\prime}s+u}{x^{\prime}s}\right),
\end{equation}
\begin{equation}
 H_{q\bar{q}}=\left(\frac{1}{6}\right)
              \left(\frac{-u}{x^{\prime}s+u}+\frac{x^{\prime}s+u}{-u}\right).
\end{equation}
It may be noticed that there are three cuts in each diagram in fig.1.
The long distance partonic distribution functions shown in eq.(15) and (16)
come from the middle cut diagrams,
and the rest part of the cut middle contribution in each diagram
combine all of the 
other cut contributions, cut left and right, to construct a  restriction
$|y^-_1| > |y^-| > |y^-_2| >0$,
which rule out the long distance effect, hence do not contribute the nuclear
enhancement.

The total contribution of the jet double scattering cross section 
with final state interaction is
\begin{eqnarray}
E_l\frac{d\sigma^{(D-f)}_{jet} (l) }{d^3l} &=&
\alpha_{em}(4\pi\alpha_s)^2\int dx^{\prime} dx\;
\delta\left(x-\frac{-x^{\prime}t}{x^{\prime}s+u}\right)\;
      \left(\frac{1}{x^{\prime}s}\right)\;
      \left(\frac{1}{x^{\prime}s+u}\right)\;
\displaystyle\sum_{q}\; e^2_q  \\ \nonumber
 & & \times
[f_{g/P}(x^{\prime})       \Phi_q(x,x^{\prime},A)H_{q} 
+f_{q/P}(x^{\prime})       \Phi_g(x,x^{\prime},A)H_{g}
+f_{\bar{q}/P}(x^{\prime}) \Phi_q(x,x^{\prime},A)H_{q\bar{q}}
],
\end{eqnarray}
where $\displaystyle\sum_q$ runs over all quark and antiquark flavors. 
The functions $\Phi_i$ with $i=q,g$ are related to the twist-4 matrix 
elements $T_i$ defined in eq.(15) and (16) as
\begin{equation}
 \Phi_i=\left[\frac{\partial^2}{\partial x^2}
        \left(\frac{T_i(x,A)}{x}\right)\right]
        \left(\frac{x^{\prime}s\;l_T}{u(x^{\prime}s+u)}\right)^2
     +  \left[\frac{\partial}{\partial x}
        \left(\frac{T_i(x,A)}{x}\right)\right] 
        \left(\frac{-x^{\prime}s}{u(x^{\prime}s+u)}\right).
\end{equation}

We adopt the same approximation for the twist-4 matrix elements
as used in ref.[4] and [5]:
\begin{equation}
    T_i(x,A)=\lambda^2 A^{4/3} f_{i/N}(x),
\end{equation}
where $i=q,\bar{q}$, and $g$, with the $f_{i/N}$ are the regular parton 
distributions in a nucleon. 
We observe that the EMC effect 
in these parton distributions can be neglected 
for the nuclear enhancement, which is pointed in ref.[6].

In this paper, we are actually working on parton double scatterings
in nuclei. We naturally assume that one initial parton 
(e.q. the $q$ in fig.3)
comes from one nucleon (e.q. the $N_a$ in fig.3)
and the other parton (e.q. the $g$ in fig.3) from another nucleon
(e.q. the $N_b$ in fig.3).
In other words, both the $g$ and $g'$ in fig.3 stand for the same gluon
from the nucleon $N_b$, hence, the distance between $g$ and $g'$
is physically zero.
Therefore, the pole structure of the propagator $A$ in fig.3
is eliminated by the zero distance between $g$ and $g'$.
Namely, this diagram, which might contribute to the cross section a little,
is not responsable for the long distance nuclear effect.
Similarly the pole structure of the propagator $B$ is also dismissed.
In this way, we can ignore the interference between the initial-state 
and final-state multiple scatterings.

In fig.4 and fig.5 we show the nuclear dependence of the final jet 
as compared with that of the direct photon within  the   
$515$ GeV proton beam on $Cu$ target. The parameter for the nuclear
dependence, $\alpha_i$ with $i=jet$ and $\gamma$,  is defined by
\begin{equation}
  A^{\alpha_i-1}\;E_l\frac{d\sigma^{(S)}_i(l)} {d^3l}
                 =E_l\frac{d\sigma^{(S)}_i(l)} {d^3l} 
                 +E_l\frac{d\sigma^{(D)}_i(l)} {d^3l}.
\end{equation}
We compare $\alpha_{jet}$ shown by solid curve with $\alpha_{\gamma}$
shown by dashed curve, and found that the jet inclusive cross section
has a large nuclear enhancement while the photon inclusion cross 
section has rather little nuclear dependence. 
It is clear that there exists an asymmetry
between the jet and the photon inclusive cross sections 
in the direct photon production process. This is because 
that the direct photon receives contribution from only initial-state
multiple scattering, while the jet are influenced by both 
the initial-state and the final-state multiple scatterings. 
In fig.4, we plot the nuclear dependence parameter $\alpha_i$ 
as a function of $x_F$ defined by $x_F=-2l_3/\sqrt{s}$ at $l_T=6$ GeV.
It is found that the asymmetry increase as $x_F$ decrease 
at fixed transverse momentum $l_T$.
In fig.5, we show the $l_T$ dependence of $\alpha_i$ at $x_F=0$.

In summary, we discussed in this paper that 
there exists an asymmetry between the jet and the photon 
inclusive cross sections 
in the reaction $P+A \rightarrow\gamma+jet+X$. 
We also show that the jet inclusive cross section 
has observable nuclear enhancement, 
which is mainly contributed from final-state multiple scatterings.


\newpage
\section*{Figure captions}

\begin{description}
\item[Fig.1.] Feynman diagrams for the three types of leading order
partonic subprocesses contributed to the final-state double scattering.
The first hard collision occurs between 
(1) gluon from proton and quark(antiquark) from nucleus.
(2) quark(antiquark) from proton and gluon from nucleus;
(3) quark(antiquark) from proton and antiquark(quark) from nucleus;

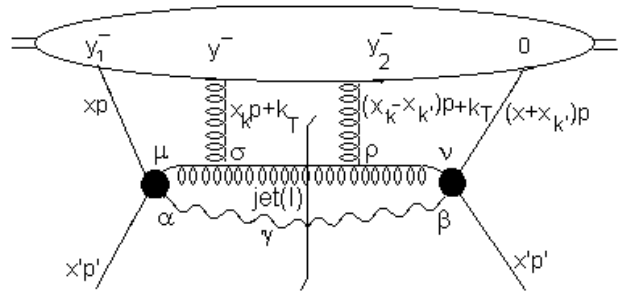
\item[Fig.2.] Graphical representation of double scattering contributions
from the parton-nucleus collisions.

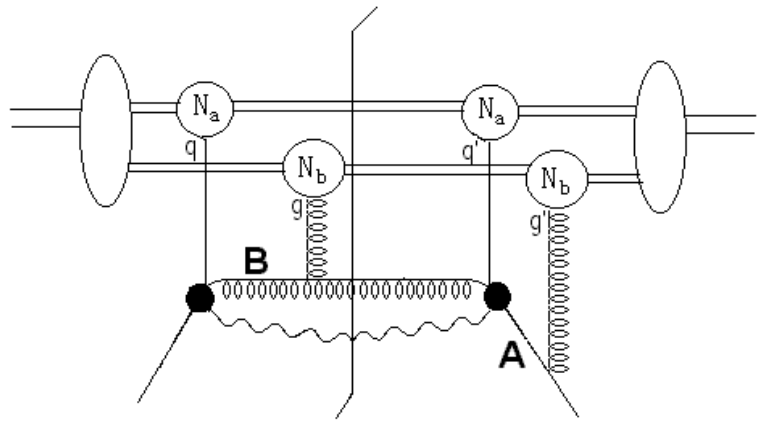
\item[Fig.3.] Graphical representation of interference between
initial-state and final-state multiple scatterings.

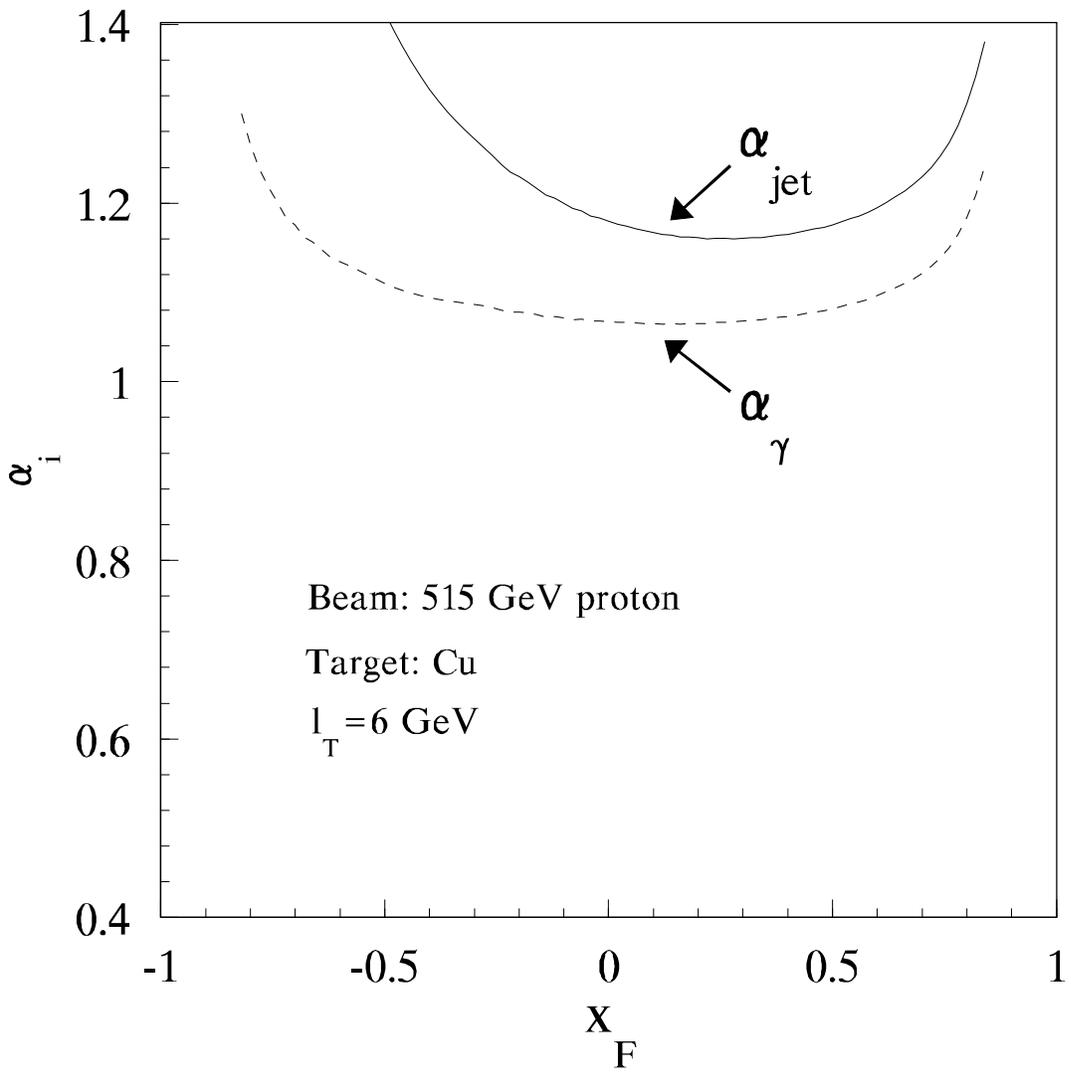
\item[Fig.4.] Behavior of $\alpha_i$, defined in eq.(26),
as a function of $x_F$ at $l_T=6$ GeV,
with $515$ GeV proton beam on $Cu$ target used. 
The  solid curve denotes the $\alpha_{jet}$, while 
the dashed curve for $\alpha_{\gamma}$.

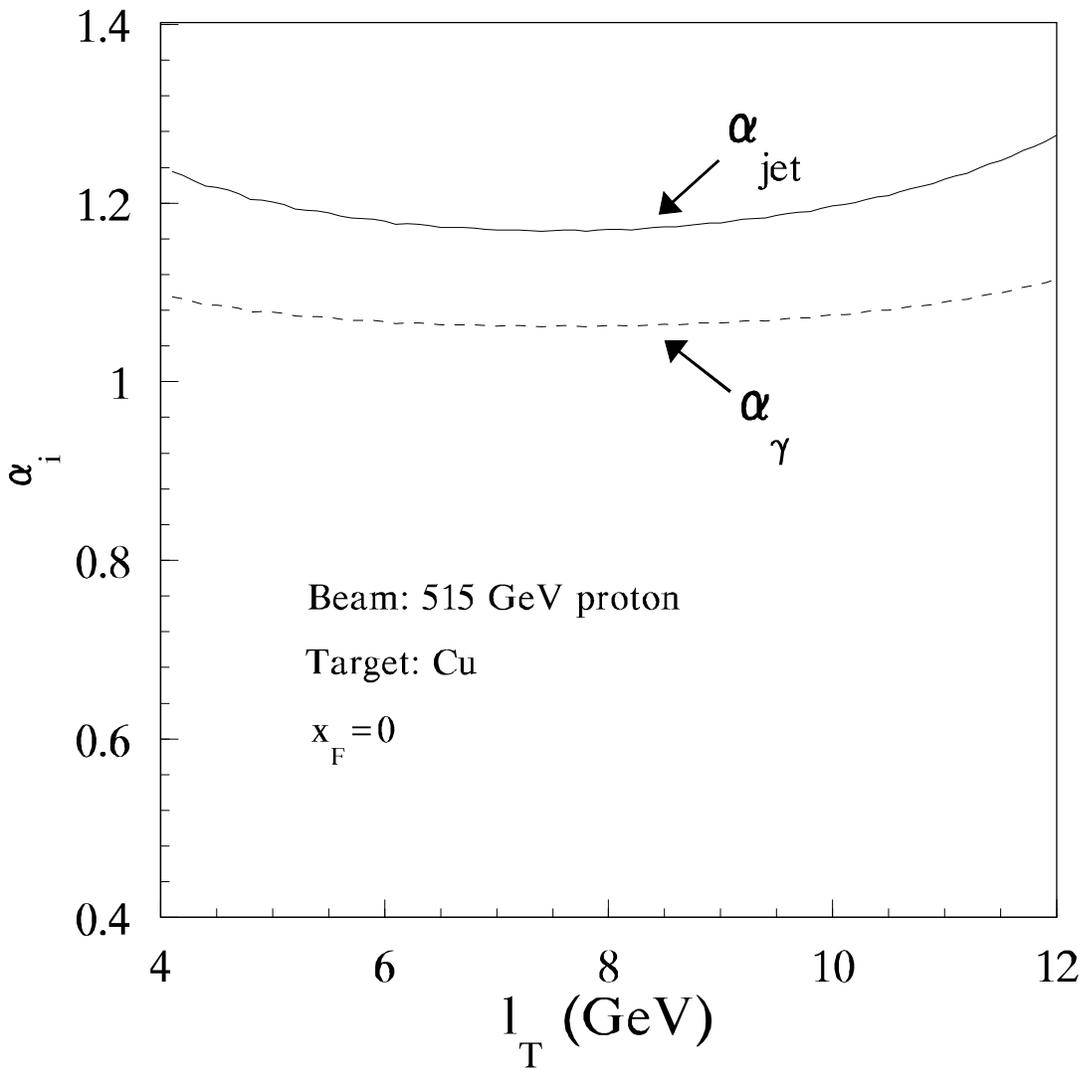
\item[Fig.5.] Behavior of $\alpha_i$, as in fig.4, but as 
a function of $l_T$ at $x_F=0$.

\end{description}
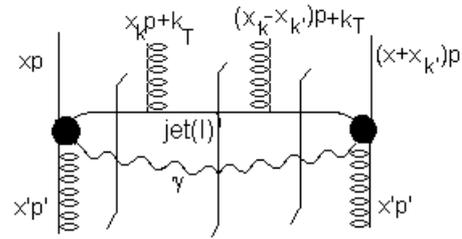
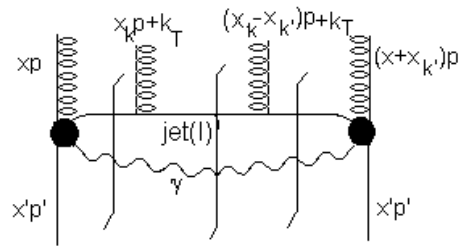
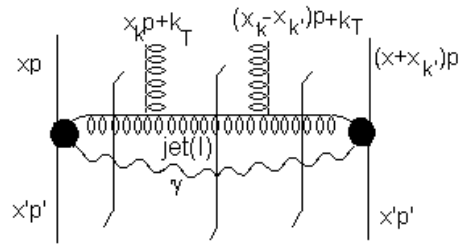
\end{document}